\documentclass[prb,twocolumn,showpacs,amsmath,amssymb,floatfix]{revtex4}

\usepackage{graphicx}
\usepackage{hyperref}

\hypersetup{%
  a4paper = {true},
  bookmarksnumbered = {true}, 
  bookmarksopen = {true}, 
  bookmarksopenlevel = {2}, 
  breaklinks = {true}, 
  colorlinks = {true}, 
  pdfauthor = {\textcopyright\ Mihal Karski, Carsten Raas, and G\"otz S. Uhrig},
  pdfcreator = {\LaTeX\ with package \flqq hyperref\frqq},
  pdfkeywords = {metal insulator transition, MIT, Hubbard model, infinite
    dimension, dynamic mean-field theory, DMFT, DMRG, D-DMRG}
  pdfmenubar = {true},
  pdfstartpage = {1},
  pdfstartview = {FitH}, 
  pdfsubject = {Electron spectra close to a metal-to-insulator transition},
  pdftitle = {Electron spectra close to a metal-to-insulator transition},
  pdftoolbar = {true},
  plainpages = {false}, 
  urlcolor = {blue} 
}

\newcommand{\Uc}{\ensuremath{U_\text{c}}}
\newcommand{\Uco}{\ensuremath{U_\text{c1}}}
\newcommand{\Uct}{\ensuremath{U_\text{c2}}}

\newif\ifadr
\adrfalse

\begin{document}

\title{Electron spectra close to a metal-to-insulator transition}

\ifadr
\else

\author{Micha{\l} Karski$^1$, Carsten Raas$^2$, and G\"otz S.~Uhrig$^2$}
\affiliation{$^1$
  Institut f\"ur Angewandte Physik der Universit\"at Bonn,
  Wegelerstr.~8,
  53115 Bonn,
  Germany}
\affiliation{$^2$
  Theoretische Physik FR 7.1,
  Geb\"aude 38,
  Universit\"at des Saarlandes,
  66123 Saarbr\"ucken,
  Germany}

\fi

\date{6 July 2005}

\begin{abstract}
  A high-resolution investigation of the electron spectra close to the
  metal-to-insulator transition in dynamic mean-field theory is presented.  An
  all-numerical, consistent confirmation of a smooth transition at zero
  temperature is provided. In particular, the separation of energy scales is
  verified. Unexpectedly, sharp peaks at the inner Hubbard band edges occur in
  the metallic regime. They are signatures of the important interaction between
  single-particle excitations and collective modes.
\end{abstract}



\maketitle



The interplay of electronic degrees of freedom with collective modes is one of
the central issues in current condensed matter physics. It is particularly
intriguing when the collective modes are formed by the electrons themselves due
to strong interactions. Famous examples for the complexity of such systems are
the high-temperature superconductors (see, e.g., Refs.~\onlinecite{norma03} and
\onlinecite{demle04}) and materials displaying colossal magnetoresistance (see,
e.g., Refs.~\onlinecite{imada98} and \onlinecite{tokur00}).

We focus on charge and spin degrees of freedom by considering a narrow
single-band model with nearest-neighbor hopping $t$ where the interaction $U$
stems from the Coulomb repulsion. For simplicity, we study the half-filled case
with one electron per site. Thus the minimal model is the Hubbard model
\begin{equation}
  \label{eq:hamilton}
  \mathcal{H} = -t \sum_{\langle i, j\rangle; \sigma}
  c^\dagger_{i;\sigma} c^{\phantom\dagger}_{j;\sigma}
  + U \sum_i (n_{i;\uparrow}-1/2) (n_{i;\downarrow}-1/2)
\end{equation}
where $i$,$j$ denote sites on a lattice with $\langle i, j\rangle$ being
nearest neighbors, $\sigma\in\{\uparrow,\downarrow\}$ the spin,
$c^{(\dagger)}_{i;\sigma}$ the electron annihilation (creation), and
$n_{i;\sigma}$ their density.

Leaving aside all effects of long-range order like charge or spin density waves
the system is metallic for weak interaction and insulating for strong
interaction. The weakly interacting system does not have significant effects of
collective modes because they are overdamped by Landau damping. The strongly
interacting system is a paramagnetic insulator governed at low energies by the
collective modes which are the magnetic excitations. The charge modes display a
large gap of the order of $U$. Thus charge and collective modes are well
separated in energy so that no significant interplay is to be expected. Hence,
the regime \emph{close} to the transition between the metal and the insulator
is the most likely to be influenced by the interplay of single-particle and
collective excitations.
 
An estimate shows that the two-dimensional superconducting cuprates are indeed
close to the metal-insulator transition. The magnetic coupling $J$ is
approximately given by $4t^2/U$; empirically, one has $J\approx t/3$ so that
$U/W\approx 1.5$ where $W=8t$ is the band width in 2D. This can be compared to
the value of about $U/W\approx1.2$ where the paramagnetic insulating phase
becomes instable due to closing of the charge gap.\cite{reisc04}

It is the aim of the present work to provide evidence that a generic
paramagnetic system with values of $U/W$ in the range $1$ to $1.5$ constitutes
a highly correlated metal with significant interplay between single-particle
and collective modes. Since a controlled treatment of finite-dimensional
systems other than the one-dimensional chain is not possible we study the
Hubbard model on the Bethe lattice with infinite coordination number
$z\to\infty$.

Scaling the hopping\cite{metzn89a} $t=D/(2\sqrt{z})$ leads to the dynamic
mean-field theory (DMFT).\cite{mulle89a,jarre92,georg92a} In DMFT the lattice
problem is mapped onto an effective single impurity Anderson model (SIAM) with
the self-consistency conditions that the interaction $U$, the full local
Green's function $G(\omega)$ \emph{and} the local self-energy $\Sigma(\omega)$
are the same in the lattice and in the SIAM.\cite{georg96} The DMFT as an
approximation to finite-dimensional systems is by now a widely employed
technique.\cite{kotli04a} It allows one to include important correlation
effects in the description of real materials based on \textit{ab initio}
density-functional theories. Hence the quantitative understanding of all
features of the DMFT solution of the Hamilton operator \eqref{eq:hamilton} is
mandatory.

At low, but finite, temperatures $T$ the phase transition between metal and
insulator is of first order\cite{georg96} and takes place at $\Uc(T)$
(Refs.~\onlinecite{tong02} and \onlinecite{blume02}) between $\Uco(T)$
(instability of the insulator against infinitesimal changes in amplitude) and
$\Uct(T)$ (instability of the metal). At $T=0$, the transition has peculiar
properties.\cite{zhang93,kotli99,bulla99,potth03b} It bears features from
first-order transitions: a jump in the entropy and a finite hysteresis between
$\Uco:=\Uco(T=0)$ and $\Uct:=\Uct(T=0)$. But there are also second-order
features because $\Uc:=\Uc(T=0)=\Uct$ where the quasiparticle weight vanishes
continuously and the ground-state energy $E(U)$ is differentiable. The behavior
of $E(U)$ is derived from a projective DMFT,\cite{moell95} which is based on
the hypothesis that the energetically high-lying spectral features do not
change at the transition from the metallic to the insulating solution
(separation of energy scales).

There are many determinations of $\Uco$ and $\Uct$. They (almost) agree on
$(2.39\pm0.02)D$ for\cite{bulla01a,blume04a,garci04} $\Uco$ with the outlier
of $(2.225\pm0.025)D$.\cite{nishi04b} For $\Uct$ the values range from $2.92D$
to $3.0D$.\cite{moell95,bulla99,bulla00,bulla01a,garci04,blume04b} The spectral
densities $\rho(\omega):=-\pi^{-1}\text{Im}\,G(\omega)$ [density of states
(DOS)] display a quasiparticle peak at $\omega=0$ in the metallic solution
which is pinned to its noninteracting value $\rho_0(0)$. But its width
decreases on $U\to \Uct$.\cite{zhang93,moell95,bulla99,garci04} The insulating
solutions display the lower and the upper Hubbard bands which merge for
$U\to\Uco$ when the single-particle gap $\Delta$
closes.\cite{zhang93,bulla01a,nishi04b,garci04}

Recent progress in the numerical calculation of dynamic quantities for quantum
impurity models\cite{gebha03,raas04a,nishi04a,raas05a} by dynamic
density-matrix renormalization\cite{hallb95,ramas97,kuhne99a} (\mbox{D-DMRG})
make calculations possible with well-controlled resolution at all energies.
Thereby, spectral functions and ground-state energies become accessible which
have so far eluded a quantitative determination. With the correction vector
method we compute $\rho(\omega)$ broadened (convolved) by Lorentzians of width
$\eta\in[0.01,0.1]D$. The unbroadened $\rho(\omega)$ is retrieved by least-bias
deconvolution.\cite{raas05a} It is used to determine the continued fraction of
the bath function in the next iteration of the DMFT self-consistency
cycle.\cite{georg96} For all $U$ about 20 iterations were performed till two
subsequent $\rho(\omega)$ differed less than $\approx10^{-3}/D$ everywhere
\emph{and} the ground-state energy and the double occupancy differ less than
$10^{-2}\%$. For the insulator, it is required in addition that the static gap,
derived from energy differences of the finite bath representation, differs less
than $1\%$.

\begin{figure}[tb]
  \centering\includegraphics[width=\columnwidth]{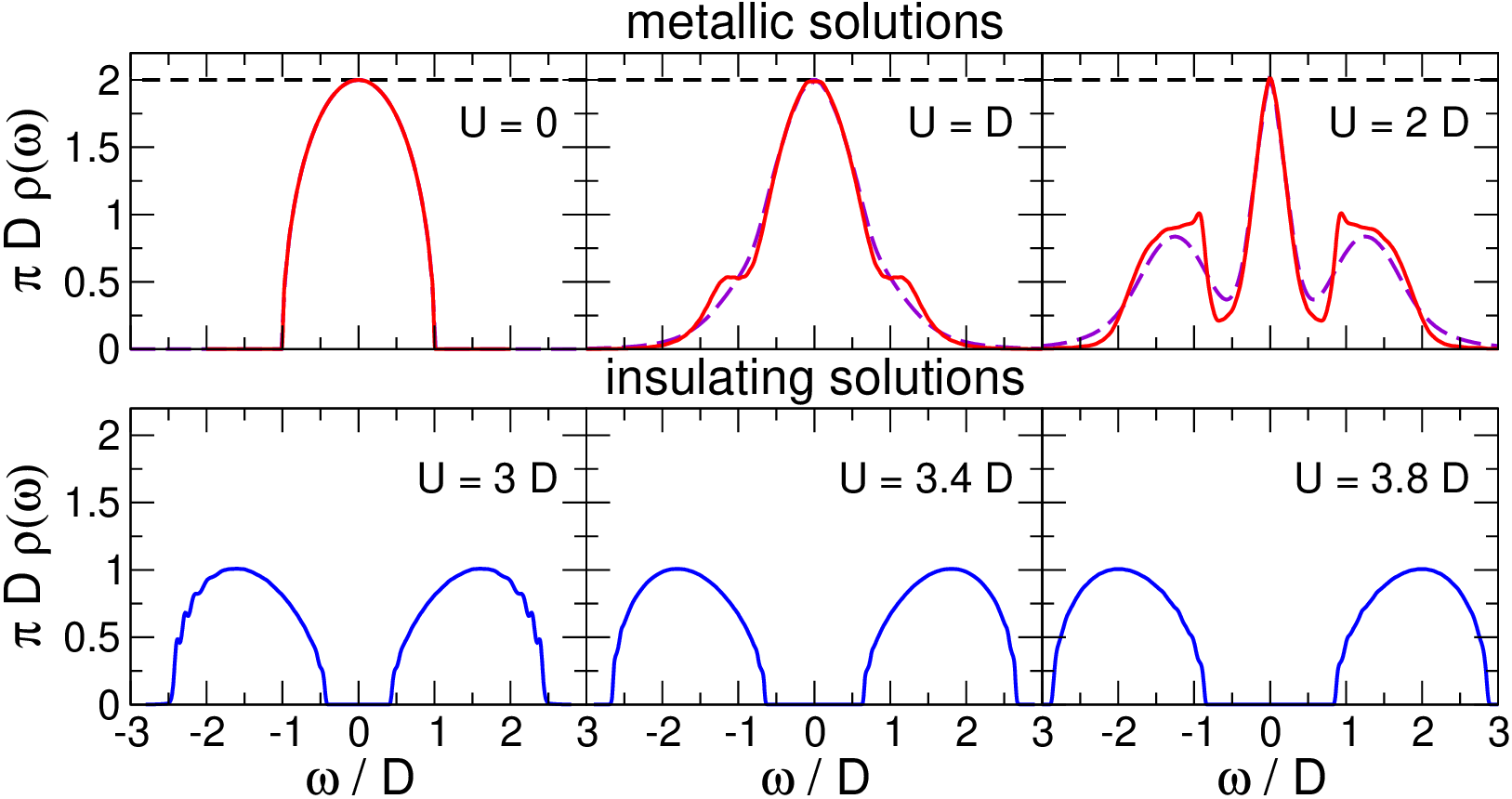}
  \caption{\label{fig:dos1}
    (Color online)
    Spectral densities $\rho(\omega)$ deep in the metallic regime (upper row)
    and deep in the insulating regime (lower row) in DMFT for the Bethe lattice
    at $T=0$; dashed lines: numerical renormalization group (NRG) data
    (Ref.~\onlinecite{bulla04p}).}
\end{figure}
The \mbox{D-DMRG} is performed with 128 or 256 basis states. We use 120, 160,
or 240 fermionic sites including the impurity in the metallic regime. For the
insulating solutions we used 121 or 161 fermions. An odd number of sites
implies a pole at $\omega=0$ in $\rho_0(\omega)$. This pole is split by the
interaction. The splitting results from a pole in $\Sigma(\omega)$ at
$\omega=0$. Such a solution is insulating. Hence an odd number of sites is
slightly biased toward an insulator. Vice versa, an even number of sites leads
to $\text{Im}\,\Sigma(0)=0$ implying a small bias toward the metallic solution.
The relative bias is estimated by the inverse number of sites:
(4--8)$\times10^{-3}$. In odd chains, we observe two almost degenerate
ground-states (spin $\uparrow$ or $\downarrow$ at the interacting site) which
must both be considered. Otherwise a spurious magnetic moment is generated.

\begin{figure}[tb]
  \centering\includegraphics[width=\columnwidth]{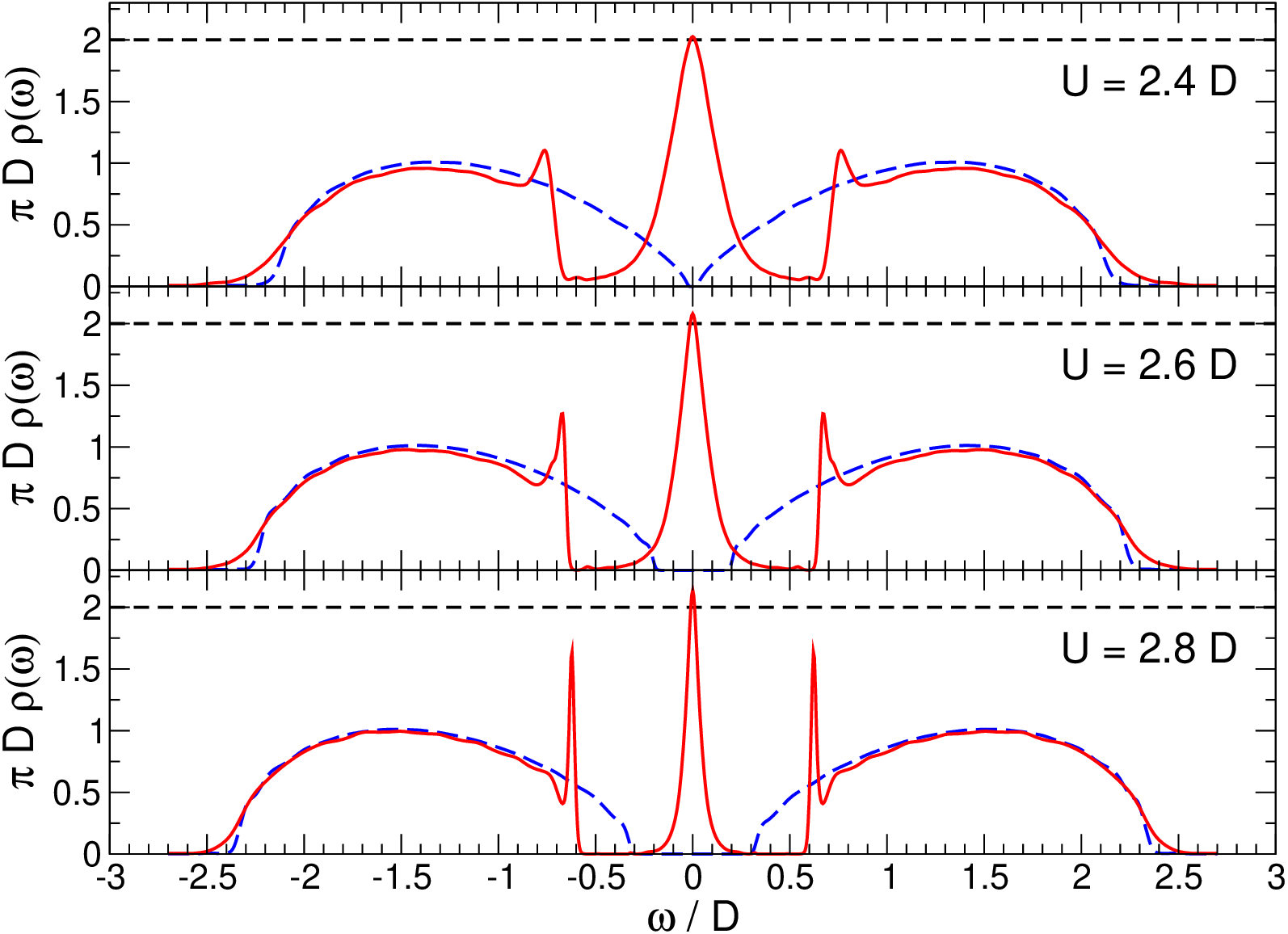}
  \caption{\label{fig:dos2}
    (Color online)
    Spectral densities $\rho(\omega)$ of the metallic (solid) and the
    insulating (dashed) solutions between $\Uco$ and $\Uct$.}
\end{figure}
In Figs.~\ref{fig:dos1} and \ref{fig:dos2}, our results for metallic and
insulating $\rho(\omega)$ are shown. In the metallic solutions, the narrowing
of the quasiparticle band around $\omega=0$ is clearly visible. From $U\approx
D$ on, the DOS displays side features which develop into the lower and upper
Hubbard bands. At $U\approx 2D$ the Hubbard bands are well separated from the
quasiparticle peak at $\omega=0$ by a precursor of the gap $\Delta$ in the
insulator: a pseudogap. The comparison with the NRG data from
Ref.~\onlinecite{bulla99} shows good agreement in the quasiparticle peak but
deviations in the Hubbard bands. There the DMRG data are much sharper and do
not have significant tails at higher energies. This difference stems from the
broadening proportional to the frequency which is inherent to the NRG
algorithm.\cite{gebha03,raas04a}

The insulating solutions display the lower and the upper Hubbard bands clearly.
They agree excellently with the perturbative result\cite{eastw03} (not shown)
for $U\gtrapprox3D$. At $U=\Uco=(2.38\pm0.02)D$ both bands touch each other. No
upturn in $\rho(\omega)$ as in Ref.~\onlinecite{nishi04b} is found when we
consider the deconvolved $\rho(\omega)$ for all $\omega$. An upturn occurs only
if the \emph{static} gap is used. But such a procedure did not lead to stable
self-consistent solutions.

\begin{figure}[tb]
  \centering\includegraphics[width=\columnwidth]{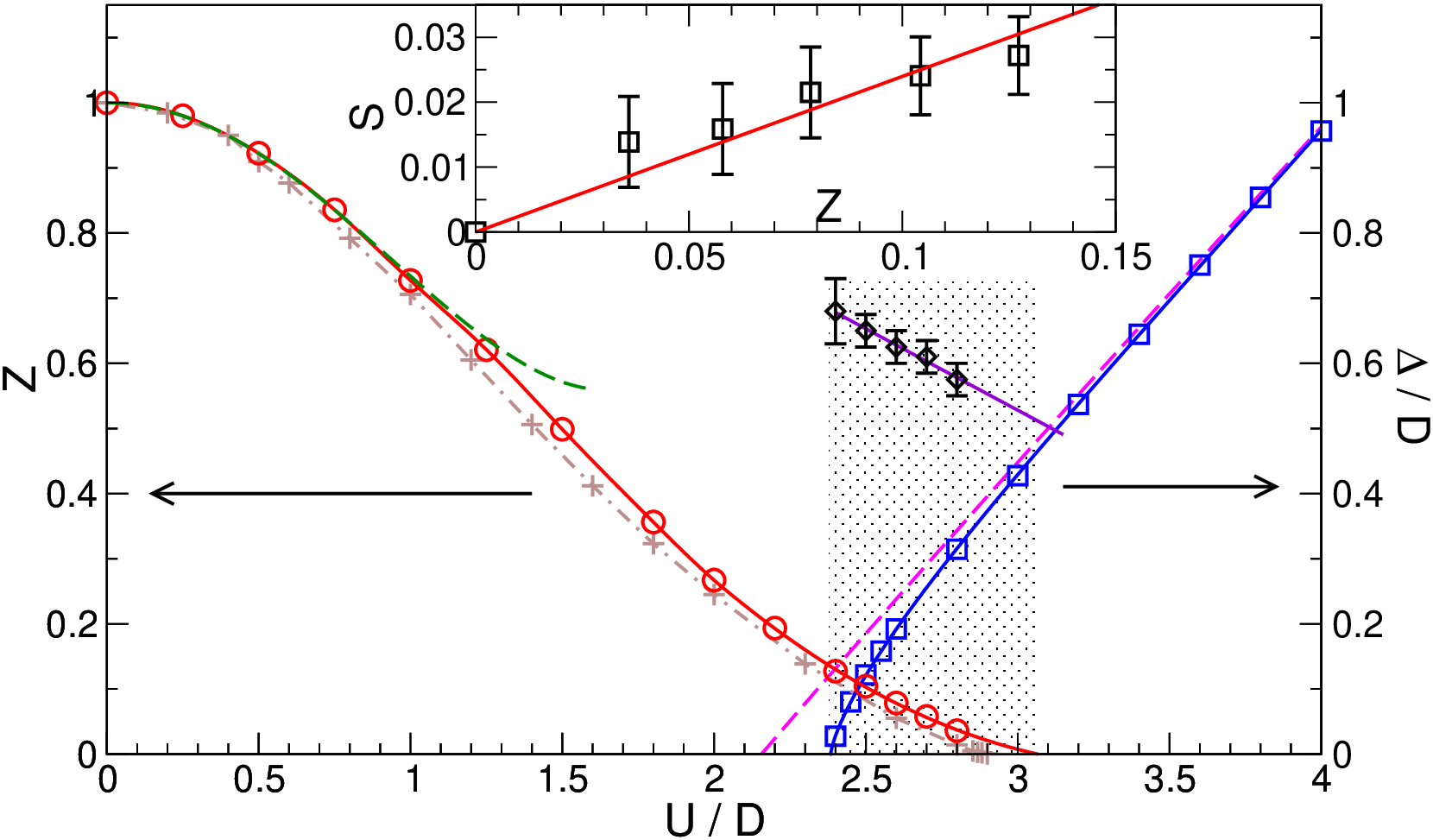}
  \caption{\label{fig:weights}
    (Color online)
    Dotted area: region of two solutions.
    Left curves: metallic quasiparticle weight $Z$;
    line with circles: interpolated DMRG,
    line with pluses: NRG\cite{bulla99};
    dashed line: perturbation up to $U^4$ (Ref.~\onlinecite{gebha03}).
    Right curves: insulating gap $\Delta$ or pseudogap in the metal
    (line with diamonds);
    line with squares: DMRG;
    dashed line: perturbation up to $1/U^2$ (Ref.~\onlinecite{eastw03}).
    Inset: weight $S$ of the peaks at the inner Hubbard band edges.}  
\end{figure}
In Fig.~\ref{fig:weights} the quasiparticle weight $Z$ in the metal and the
single-particle gap $\Delta$ in the insulator are shown. The weight
$Z = [1-\partial_\omega\text{Re}\,\Sigma(0)]^{-1}$ is found from fitting the
derivative of the Dyson equation $G(\omega) = G_0[\omega-\Sigma(\omega)]$
implying $Z^{-1} = D^2\partial_\omega G(0)/2$ where $G_0(\omega)$ is the bare
local Green's function of the lattice. The gap $\Delta$ is found from a fit
proportional to $\sqrt{\omega-\Delta}$ to the sharp uprise of the DOS at the
inner band edges. The DMRG data agree excellently with the perturbative
results where the respective perturbation holds. The comparison to the NRG data
shows that the broader high energy features reduce $Z$ to some extent so that
the NRG weight stays below the DMRG data.

From the power-law fit $\Delta = (U-\Uco)^\zeta[a_1+a_2(U-\Uco)]$ shown in
Fig.~\ref{fig:weights} we find $\Uco=(2.38\pm0.02)D$ in perfect agreement with
most of the previous results.\cite{bulla01a,blume04a,garci04} The exponent is
found to be $\zeta=0.72\pm0.05$. The value for $\Uct=(3.07\pm0.1)D$ is
determined reliably from a second-order extrapolation shown in
Fig.~\ref{fig:weights}. The value of $\Uct$ agrees well with the previous
results.\cite{moell95,bulla99,bulla00,bulla01a,garci04,blume04b} We attribute
the small deviation to the enhanced accuracy of our \mbox{D-DMRG} algorithm.

Figure~\ref{fig:dos2} nicely shows the evolution of the metallic and the
insulating $\rho(\omega)$ between $\Uco$ and $\Uct$. It represents the first
all-numerical high-precision confirmation of the hypothesis of the separation
of energy scales. Clearly, the metallic Hubbard bands at higher energies
approach the insulating ones for $U\to\Uct$. Differences remain at the inner
band edges and around $\omega=0$ as long as $U<\Uct$. The frequency of the
sharp uprise in $\rho(\omega)$ at the inner edges in the metallic solutions
defines the pseudogap. Its evolution (line with diamonds in
Fig.~\ref{fig:weights}) shows that it tends to the insulating gap at $U=\Uct$.
This corroborates strongly that the metallic $\rho(\omega)$ approaches the
insulating $\rho(\omega)$ for $U\to\Uct$. Remarkably, the metal and the
insulator have the same single-particle correlations at $U=\Uct$, which is
again a feature of a second-order phase transition.

\begin{figure}[tb]
 \centering\includegraphics[width=\columnwidth]{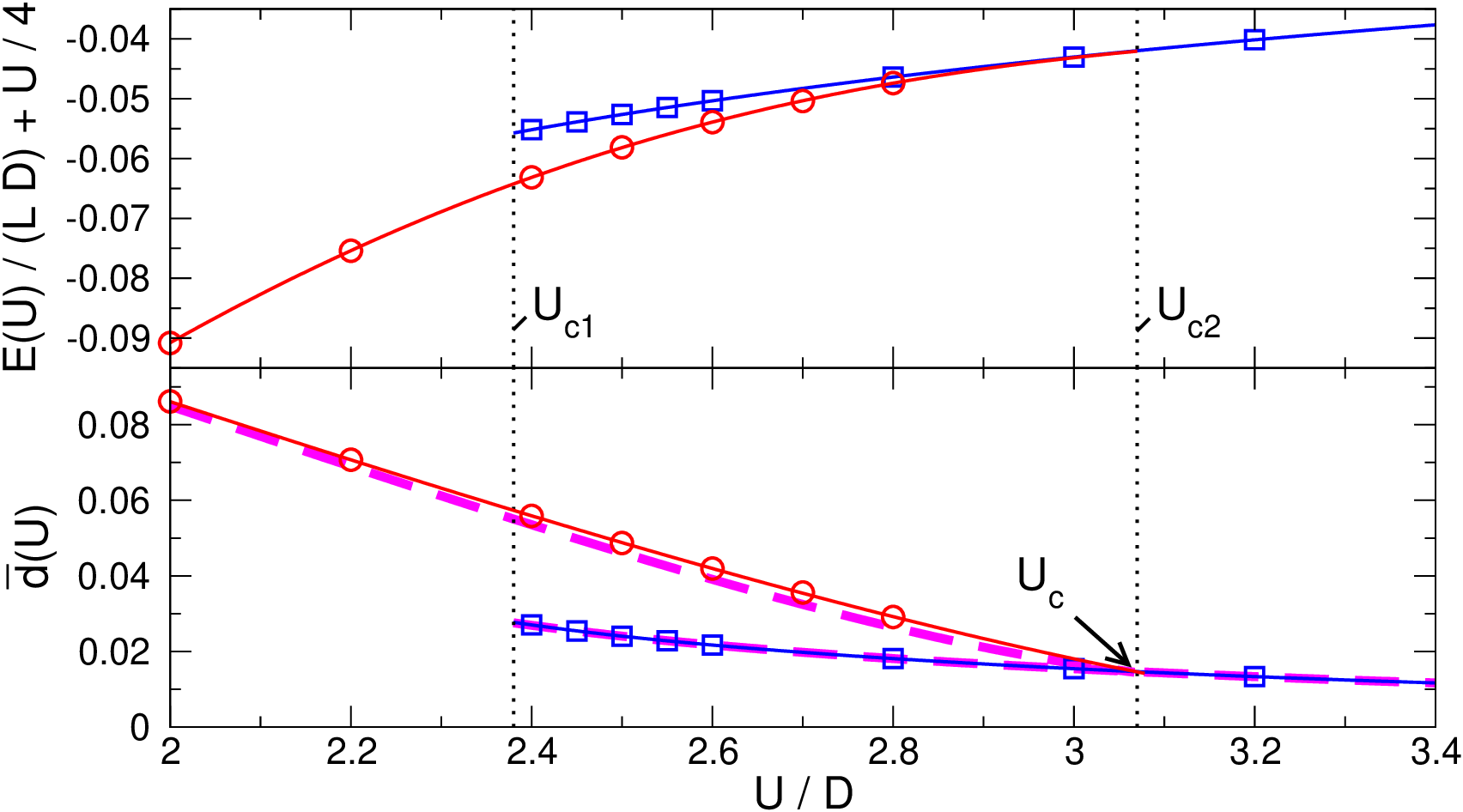}
 \caption{\label{fig:energy}
   (Color online)
   Upper panel: ground-state energy $E(U)$ in the insulator (squares) and in
   the metal (circles).
   Lower panel: corresponding double occupancy $\overline{d}(U)$.
   At $\Uc=(3.07\pm0.04)D$ the double occupancies intersect.
   The dashed lines depict the derivative of the interpolations for $E(U)$.}
\end{figure}
From equations of motion\cite{nishi04b} one obtains the ground-state energy
per site
\begin{equation}
  \frac{E(U)}{L} = \sum_\sigma 
  \left\langle
    c^\dagger_{i;\sigma}[ c^{\phantom\dagger}_{i;\sigma},\mathcal{H}]
  \right\rangle
  - U \left(\overline{d}(U)-\frac{1}{4}\right)
\end{equation}
where $\overline{d}(U)$ is the double occupancy $\langle n_{i;\uparrow}
n_{i;\downarrow}\rangle$. All the local expectation values are computed
directly at the interaction site of the SIAM at self-consistency. Then they
equal the local expectation values on the lattice. In Fig.~\ref{fig:energy},
$E(U)$ and $\overline{d}(U)$ are depicted. The equality of $\overline{d}(U)$ in
the SIAM [data points and interpolations (solid lines)] and in the lattice,
obtained from $\overline{d}(U) = \partial_U E+1/4$ (dashed lines), is a very
sensitive check of the data quality. It is perfectly fulfilled in the insulator
(difference $\lessapprox2\times10^{-4}$). The comparison of energies after
integrating $\overline{d}(U)$ for $U\in[2.4,4.0]D$ as in
Ref.~\onlinecite{nishi04b} shows agreement within $8\times10^{-6}D$. In the
metal, our check for $\overline{d}(U)$ works very well, but not perfectly
(difference $\lessapprox3\times10^{-3}$). The remaining deviation is a
finite-size effect of the bath because the deviation decreases on increasing
the number of fermionic sites.
 
Assuming a differentiable behavior of $E(U)$ we deduce the critical interaction
$\Uc=(3.07\pm0.04)D$ from the intersection of the double occupancies. The
corresponding metallic and insulating energies differ less than $10^{-4}D$.
This justifies the assumption of differentiability \textit{a posteriori}. The
agreement of $\Uc$ and $\Uct$ in our numerical treatment proves the consistency
of our data. Thereby, the previously proposed
scenario\cite{zhang93,moell95,kotli99} for the metal-insulator transition is
numerically confirmed.

An unexpected feature in our metallic solution are the sharp peaks at the inner
edges of the Hubbard bands (see Fig.~\ref{fig:dos2} and Fig.~\ref{fig:dos1} for
$U=2D$). The only previous evidence for similar features were weak shoulders in
NRG\cite{bulla99} and quantum Monte Carlo\cite{blume02} data. Based on the
following arguments we interpret it as the signature of a collective mode.

Since the sharp peaks occur only in the metal the heavy quasiparticles must be
involved. But the peaks are located at relatively high energies compared to the
quasiparticle band. So a binding or antibinding phenomenon with something else
must take place. Since the peaks are very sharp (to numerical accuracy for
$U\gtrapprox 2.6D$) we conclude that an immobile, local mode is involved. Spin
excitations like spin waves are the best-known collective excitations in
Hubbard systems. In infinite dimensions, they are indeed
dispersionless.\cite{klein95} In absence of any magnetic order they are located
at zero energy in the insulator because the Hartree term
vanishes.\cite{mulle89a} By continuity, we deduce that their energy is very low
also in the metal with an estimated upper bound of $0.2D$ deduced from the
energy of a spin wave in the long-range ordered N\'eel state which is $z
J/2\approx D^2/(2U)\approx0.2D$ at $U=2.5D$. Hence, we conclude from the energy
of the sharp features that it signals the \emph{antibound} state or resonance
of a heavy quasiparticle with a collective spin excitation. We refer to this
state as the antipolaron.

In the inset of Fig.~\ref{fig:weights} we plot the weight $S$ of the
antipolaron peak as a function of $Z$. The error bars result from the numerical
difficulty to resolve this sharp feature and from the analytical difficulty to
separate it from the background of the Hubbard bands. For the separation, we
fitted an approximate square root onset starting at the pseudogap, multiplied
with a quadratic polynomial, to the Hubbard bands. The excess weight $S$ is
attributed to the antipolaron peak. From the inset in Fig.~\ref{fig:weights} we
conclude that $S$ vanishes linearly with $Z$, rather than quadratically or
cubically. Thus the antipolaron peak vanishes proportional to the matrix
element of a \emph{single} quasiparticle. If $S$ had vanished quadratically
(cubically) one would have concluded that two (three) quasiparticles were
involved. The linear dependence of $S$ on $Z$ corroborates the scenario of an
antibound state formed from one quasiparticle and one collective mode. Surely,
further investigations are called for.

We point out that the antibinding between the heavy quasiparticles and the
collective magnetic modes suggests an interesting answer to the so far open
question why the metal forms Hubbard bands and eventually becomes insulating.
The weight close to the Fermi level is pushed away to higher energies by a
strong repulsive interaction between the low-lying quasiparticles and magnetic
modes. Note that this scenario is also applicable in finite dimensions if the
finite dispersion of the collective modes is accounted for.

In summary, we provided the all-numerical confirmation of the metal-insulator
transition proposed earlier\cite{zhang93,moell95,kotli99} on the basis of the
hypothesis of the separation of energy scales. We found that the spectral
density of the metal approaches the one of the insulator for $U\to\Uct$. The
critical interactions $\Uco=(2.38\pm0.02)D$ and $\Uct=(3.07\pm0.1)D$ were
found; the latter value coincides with $\Uc=(3.07\pm0.04)D$ where the
ground-state energies $E(U)$ intersect differentiably. Hence, a consistent
picture of a differentiable $E(U)$ emerged.

The unprecedented resolution of the spectral densities enabled us to detect
sharp peaks at the inner Hubbard band edges. We interpreted them as antibound
states or resonances formed from a heavy quasiparticle and a collective
magnetic mode. The occurrence of such signatures in electron spectra opens up
an interesting route to investigate the interplay of single-particle and
collective excitations in photoelectron spectroscopy.


\mbox{}

  We thank R.~Bulla for the NRG data and E.~M\"uller-Hartmann, A.~Rosch, and
  L.~H.~Tjeng for helpful discussions and the DFG for financial support in
  SFB~608.



\end{document}